\documentclass{article}

\def\mytitle#1{\setcounter{equation}{0}
\setcounter{footnote}{0}
\begin{flushleft}\Large\textbf{#1}\end{flushleft}
\vspace{0.25cm}}
\def\myname#1{\leftline{{\large #1}}\vspace{-0.13cm}}
\def\myplace#1#2{\small\begin{flushleft}\textit{#1}\\
\texttt{#2}\end{flushleft}}

\usepackage{graphicx}% Include figure files
\begin{document}

\mytitle{HOW DOES INFLATION DEPEND UPON THE NATURE OF FLUIDS FILLING UP THE UNIVERSE IN BRANE WORLD SCENARIO}

\myname{ Sudeshna Mukerji~$*$} \vskip0.2cm \myname{ Ritabrata
Biswas $*$} \vskip0.2cm\myname{Nairwita Mazumder$*$}
\vskip0.2cm\myname{Subenoy Chakraborty$*$}

\myplace{Department of Mathematics, Jadavpur University,
Kolkata-32, India.}
{[mukerjisudeshna@gmail.com,biswas.ritabrata@gmail.com~,~nairwita15@gmail.com,schakraborty@math.jdvu.ac.in]}

%\myclassification{}

\begin{abstract}

By constructing different parameters which are able to give us the information about our universe during inflation,(specially at the start and the end of the inflationary universe) a brief idea of brane world inflation is given in this work. What will be the size of the universe at the end of inflation,i.e.,how many times will it grow than the original size is been speculated and analysed thereafter. Different kinds of fluids are taken to be the matter inside the brane. It is observed that in the case of highly positive pressure grower gas like polytropic,the size of the universe at the end of inflation is comparitively smaller. Whereas for negative pressure creators (like Chaplygin gas) this size is much bigger. Except these two cases, inflation has been studied for barotropic fluid and linear red shift parametrization $\omega(z) = \omega_{0} + \omega_{1} z$ too. For them the size of the universe after inflation is much more high. We also have seen that this size does not depend upon the potential energy at the end of the inflation. On the contrary, there is a high impact of the initial potential energy upon the size of inflation.
\end{abstract}

\section{Introduction}
Inflation is the most promising candidate for understanding the physics of the very early universe, typing the evolution of the universe to the properties of one or more scalar inflaton fields, responsible for creating an accelerating expanding universe. This then creates a flat and homogeneous universe which later evolves into the present universe. The accelerated expansion of the universe has now been well documented in the literature \cite{Perlmutter1} and is strongly confirmed by the cosmic microwave back ground radiation(CMBR) \cite{Spergel1, Spergel2} and Sloan Digital Sky Survey \cite{Einstein1}. Inflationary scenario resolves many puzzles of the standard Hot Big Bang Cosmology like the horizon and flatness problems. Another success of the inflatioanary universe model is that it provides a causal interpretation of the origin of the observed anisotropy of the CMBR and also the distribution of large scale structures \cite{delCampo1, Guth1, Albrecht1, Dunkley1, Hinshaw1}.

Recently, there has been a proliferation of brane world models, where the standard model matter is confined in a four dimensional space time said to be a singular hyper surface or 3-brane embedded in a $(4+d)$-dimensional space time called the bulk. Though the matter is all confined to a 3-brane (related to open string modes), the gravitational field can propagate in the vacuum filled five dimensional bulk (closed string mode). The effect of the extra dimension induces additional terms in the Friedmann equation \cite{Binetrug1, Shiromizu1}. One of the term that appears in this equation is a quadratic term in the energy density. Such a term generally makes it easier to obtain inflation in the early universe \cite{Mohapatra1, Maartens1}.

It is important to see if the standard expanding universe can be recovered by extending the static solution to a time-dependent one when matter/radiation inside the brane is included. As shown in \cite{Csaki1}, the standard matter-dominated expanding universe is recovered for large enough
brane tension. However, \cite{Csaki1} proposed two possible behaviors for the cosmic scale factor,
a(t), during the radiation-dominated universe: (i) $a(t) \sim t^{\frac{1}{4}}$ (first found in \cite{Bin1}), or (ii)
$a(t) \sim t^{\frac{1}{2}}$ (found in \cite{Csaki1}). That there might be two different powerlaw solutions is not
surprising, since a non-linear second order differential equation may have more than one
such solution, and initial conditions will determine the precise time evolution of $a(t)$ (which
may not be exactly a powerlaw, but might be close to either of the candidate powerlaws
at different times). On the other hand, to match the known observations of the expanding
Universe, at least back to the time of electron-positron annihilation and nucleosynthesis, the
expansion rate of the Universe should be approximately $H^{2} = \frac{8\pi G \rho}{3}$, its value in standard
Big Bang cosmology. Actually this requirement is stronger than demanding $a(t) \sim t^{\frac{1}{2}}$ :
agreement with the well-established picture of light element synthesis in the early Universe
\cite{Peebles1} constrains H as a function of temperature T for $T \sim 100$ keV.

The four dimensional Einstein equations induced on the brane can be written as \cite{Shiromizu1}
\begin{equation}\label{1}
G_{\mu\nu}=-\Lambda_{4}g_{\mu\nu}+\left(\frac{8\pi}{M_{p}^{2}}\right)T_{\mu\nu}+\left(\frac{8\pi}{M_{5}^{3}}\right)^{2}\pi_{\mu\nu}-E_{\mu\nu}
\end{equation}
where $T_{\mu\nu}$ is the energy momentum tensor of matter on the brane, $\pi_{\mu\nu}$ is a tensor quadratic in $T_{\mu\nu}$, and $E_{\mu\nu}$ is a projection of the five dimensional Weyl tensor. The effective cosmological constant $\Lambda_{4}$ on the brane is given by
\begin{equation}\label{2}
\Lambda_{4}=\frac{4\pi}{M_{5}^{3}}\left(\Lambda+\frac{4\pi}{3M_{5}^{3}}\sigma^{2}\right)
\end{equation}
where $\Lambda$ is the five dimensional bulk cosmological constant and $\sigma$ is the 3-brane tension.

The four dimensional Plank scale is given by
\begin{equation}\label{3}
M_{p}=\sqrt{\frac{3}{4\pi}}\left(\frac{M_{5}^{2}}{\sqrt{\sigma}}\right)M_{5}
\end{equation}
The Friedmann equation on the brane has the generalised form \cite{Binetroy, Flanagan}
\begin{equation}\label{4}
H^{2}=\left(\frac{8\pi}{3M_{p}^{2}}\right)\rho+\left(\frac{4\pi}{3M_{5}^{3}}\right)\rho^{2}+\frac{\Lambda_{4}}{3}+\frac{\xi}{a^{4}}
\end{equation}
where $\xi$ is an integration constant  arising from $E_{\mu\nu}$ and thus transmitting bulk gravitational influence onto the brane and $a(t)$ is the scale factor and $H=\frac{\dot{a}}{a}$ denotes the Hubble parameter and $\rho$ represents the energy density of matter.

We assume that the four dimensional cosmological constant $\Lambda_{4}$ is set to zero and once inflation begins the final term will rapidly become unimportant leaving us with \cite{Maartens1, Herrera1}
\begin{equation}\label{5}
H^{2}=\kappa f\left(\rho_{\phi}\right)\left[1+\frac{f\left(\rho_{\phi}\right)}{2\sigma}\right]
\end{equation}
where $\kappa=\frac{8\pi G}{3}=\frac{8\pi}{3M_{p}^{2}}$
and $\rho_{\phi}$ represents the matter confined to the brane.

Here, $\rho_{\phi}$ becomes $\rho_{\phi}=\frac{\dot{\phi}^{2}}{2}+V(\phi)$ and $V(\phi)=V$ is the scalar potential.

We assume that the scalar field is confined to the brane, so that its field equation has the standard form
\begin{equation}\label{6}
\ddot{\phi}+3H\dot{\phi}+V'=0
\end{equation}
where dots mean derivatives w.r.t. the cosmological time and $V'=\frac{\partial V(\phi)}{\partial \phi}$. For convinience we use units in which $c=\hbar=1$.

Also, we introduce the density matter $\rho_{m}\sim a^{-3}$ in such a way using extrapolation that we may write \cite{Herrera1}
\begin{equation}\label{7}
\rho^{*}=f\left(\rho_{m}\right)\longrightarrow f\left(\rho_{\phi}\right)
\end{equation}
where $\rho^{*}$ is the energy density of barotropic fluid, or polytropic fluid or modified Chaplygin Gas(MCG hereafter) or fluid having variable equation of state is discussed in the subsequent sections. We identify $\rho_{m}$
with the contributions of the scalar field which gives equation (\ref{5}).

During the inflationary epoch the energy density associated to the scalar field is of the order of the potential, i.e., $\rho_{\phi}\sim V$.
In most of the studies about inflation, the field $\phi$ is assumed to be "slowly rolling" during most of the inflationary epoch. Physically slow roll approximation means that magnitude $\ddot{\phi}$ is irrelevant to constant multiplied by the $\dot{\phi}$ as well as $\frac{dV}{d\phi}$, or that potential must be sufficiently flat enough (small derivatives) so that the field rolls slowly enough for inflation to occur. The slowly rolling approximation means that the motion of the inflation field is overdamped, $\ddot{\phi}=0$ so that equation  $$\ddot{\phi}+\left(3H +\Gamma\right)\dot{\phi}+\frac{dV}{d\phi}=0$$ becomes a frst order equation, the $\Gamma\dot{\phi}$ term is also generally negligible during this part of inflation. The motion is controlled entirely by the force term $\frac{dV}{d\phi}$ and the viscous damping term ($3H\dot{\phi}$) due to the expansion of the universe. Near the end of the inflationary epoch, the field approaches the minimum of the potential (i.e., true vacuum) and then oscilates about it , while the term $\Gamma\dot{\phi}$ gives rise to particle and entropy production In this manner, a "graceful exit" to inflaton is achieved.
Assuming the set of slow roll conditions, i.e., $\dot{\phi}^{2}\ll V(\phi)$ and $\ddot{\phi}\ll 3H\dot{\phi}$, the Friedmann equation (\ref{5}) reduces to
\begin{equation}\label{8}
H^{2}\approx\kappa f(V)\left[1+\frac{f(V)}{2\sigma}\right]
\end{equation}
and equation (\ref{6}) becomes
\begin{equation}\label{9}
3H\dot{\phi}\approx -V'
\end{equation}
Introducing the dimensionless slow roll parameters \cite{Herrera1, Maartens2}, we write
\begin{equation}\label{10}
\epsilon=-\frac{\dot{H}}{H^{2}}
\end{equation}
and
\begin{equation}\label{11}
\eta=-\frac{\ddot{\phi}}{H \dot{\phi}}\simeq \frac{V''}{3H^{2}}
\end{equation}
where $max\left\{\epsilon, \left|\eta\right|\right\}\ll 1$.

The condition for inflation is
\begin{equation}\label{12}
\ddot{a}>0~~~~~ i.e.,~~~~~ \epsilon<1
\end{equation}
Inflation ends when the universe starts heating at a time when
\begin{equation}\label{13}
\epsilon\simeq 1.
\end{equation}

The number of e-folds during inflation is given by $N=\int_{t_{i}}^{t_{f}}Hdt$. This number is an important quantity which indicates the proper multiplication of the size of the universe as the inflation ends to get rid of the problems encountered by a standard big bang model. Using slow roll approximation it becomes
\begin{equation}\label{14}
N=-\frac{8\pi}{M_{p}^{2}}\int_{\phi_{i}}^{\phi_{f}}\frac{H^{2}}{V'}d\phi
\end{equation}
or equivalently,
\begin{equation}\label{15}
N=-\frac{8\pi}{M_{p}^{2}}\int_{V_{i}}^{V_{f}}\frac{H^{2}}{V^{'2}}dV
\end{equation}
where $t_{i}$, $\phi_{i}$, $V_{i}$ and $t_{f}$, $\phi_{f}$, $V_{f}$ are the values of the cosmological time $t$, inflaton scalar field $\phi$ and the potential V at the beginning and the end of the inflation respectively.

In this paper we wish to find out the condition for inflation and the number of e folds during inflation for a suitable choice of chaotic potential and analyse these situations graphically.
In section 2 we will study the inflation when the universe is filled up by fluid of barotropic nature. In section 3 and section 4 the fluid types are taken to be of polytropic and MCG type. In section 5 we will consider the universe filled by the fluid obeying the equation of state(EoS hereafter) $p=\omega\rho$ where varying EoS parameter is chosen as two index parametrization model namely :linear red shift parametrization  : $\omega(z)=\omega_{0}+\omega_{1}z$  where $\omega_{0}$ and $\omega_{1}$ are constants, z is the red shift parameter. Finally in section 6 a brief conclusion will be given.

\section{Inflation with barotropic fluid on the brane}
In this section we will take the fluid  which is present as the matter confined in the brane be of barotropic nature for which the pressure will be directly proportional to the fluid density. Barotropic fluids are also important idealized fluids in astrophysics, such as in the study of stellar interiors or of the interstellar medium.

The energy density and pressure are related by a
barotropic equation of state
\begin{equation}\label{16}
p = \left(\Gamma - 1\right)\rho,~~~~~~~~~ 0 \leq \Gamma \leq 2.
\end{equation}
Here $p$ and $\rho$ are the pressure and the density of the barotropic fluid respectively and $A=\left(\Gamma - 1\right)$ being the proportioality constant where
 $\Gamma=\frac{4}{3}$
corresponds to a pure radiation, $\Gamma=1$  to a pressureless dust,
to an inflationary fluid, $\Gamma=0$  to a phenomenological matter sector
cosmological constant, and $\Gamma=2$  to a stiff fluid.

The energy conservation equation is given by,
\begin{equation}\label{17}
\dot{\rho}+3\left(\rho+p\right)\frac{\dot{a}}{a}=0
\end{equation}

Now using (\ref{16}) in (\ref{17}) and solving the differential equation we have,
\begin{equation}\label{18}
\rho^{*}=\frac{\rho_{0}}{a^{3\left(A+1\right)}}.
\end{equation}
Here, $\rho_{0}=$ the positive constant of integration. We can easily see that $\rho_{0}=\rho^{*}$ when $z=0$, i.e., $\rho_{0}$ denotes the density of the universe when its voume is constant.

Now, during the inflationary epoch equation (\ref{5}) takes the form
\begin{equation}\label{19}
H^{2}=\kappa V^{A+1}\left[1+\frac{V^{A+1}}{2\sigma}\right]
\end{equation}
So from equations (\ref{10}) and (\ref{11}) we obtain the dimensionless slow parameters
\begin{equation}\label{20}
\epsilon \simeq \frac{M_{p}^{2}\left(A+1\right)V^{A} V'^{2}\left[1+\frac{V^{A+1}}{\sigma}\right]}{16 \pi V^{2\left(A+1\right)}\left[1+\frac{V^{A+1}}{2\sigma}\right]^{2}}
\end{equation}
and
\begin{equation}\label{21}
\eta \simeq \frac{M_{p}^{2}V''}{8 \pi V^{\left(A+1\right)}\left[1+\frac{V^{A+1}}{2\sigma}\right]}
\end{equation}
Now the condition (\ref{12}) leads us to the inequality
\begin{equation}\label{22}
V^{A}V'^{2}\left[1+\frac{V^{A+1}}{\sigma}\right]<\frac{16\pi}{M_{p}^{2}(A+1)}V^{2(A+1)}\left[1+\frac{V^{A+1}}{2\sigma}\right]^{2}
\end{equation}
Inflation ends when the universe starts heating at a time when $\epsilon \simeq 1$ implying
\begin{equation}\label{23}
V_{f}^{A}V_{f}'^{2}\left[1+\frac{V_{f}^{A+1}}{\sigma}\right]^{2} \simeq \frac{16\pi}{M_{p}^{2}(A+1)}V_{f}^{2(A+1)}\left[1+\frac{V_{f}^{A+1}}{2\sigma}\right]^{2}
\end{equation}
Now in the high energy limit, $V_{f}^{A+1}\gg \sigma$
and equation (\ref{23}) reduces to

$$V_{f}^{'2}\simeq \frac{\frac{16\pi}{M_{p}^{2}}\frac{1}{A+1}V_{f}^{2(A+1)}\left[1+\frac{V_{f}^{A+1}}{\sigma}+\frac{V_{f}^{2(A+1)}}{4\sigma^{2}}\right]}{V_{f}^{A}\left[1+\frac{V_{f}^{A+1}}{\sigma}\right]}$$
\begin{equation}\label{24}
=\frac{4\pi V_{f}^{3(A+1)}}{\sigma M_{p}^{2}(A+1)V_{f}^{A}}~~~~~~~~~~~~~~~~~~~~~~~~~~
\end{equation}
The number of e-folds during inflation is given by
\begin{equation}\label{25}
N=-\frac{8\pi}{M_{p}^{2}}\int_{V_{i}}^{V_{f}}\frac{1}{V'^{2}}\left[V^{A+1}\left(1+\frac{V^{A+1}}{2\sigma}\right)\right]dV
\end{equation}
In the high energy limit $\rho_{\phi}^{A+1}\approx V^{A+1}\gg \sigma$ and so
\begin{equation}\label{26}
N=-\frac{4\pi}{\sigma M_{p}^{2}}\int_{V_{i}}^{V_{f}}\frac{V^{2(A+1)}}{V'^{2}}dV
\end{equation}
\subsection{Standard quadratic Scalar Potential(SSP)}
The simplest chaotic inflation model is that of a free field with a quadratic potential,$V=\frac{1}{2}m^{2}\phi^{2}$
, where $m$ represents the mass of the inflaton field. Now using this potential in (\ref{26}) we get,
\begin{equation}\label{27}
N=\frac{\pi 4^{A}}{\sigma M_{p}^{2} (A+1) m^{2}}\left[V_{i}^{2(A+1)}-V_{f}^{2(A+1)}\right]
\end{equation}

In figure 1a we have plotted the variation of $N$ with the change of $V_{f}$ and $A$. We have given all the other parameters(upon which the N is depending) standard values. The initial potential energy of the universe must be larger than the final as after inflation when the universe has consumed a larger volume then we can assume that a reasonably larger percentage of the potential energy has turned into dynamical energy. So here we have taken $V_{f}$ hundred times smaller than the inital potential energy. It is to be noted that the number of e-folds has no remarkable change with the change of the $V_{f}$. But as we increase the value of A in positive value zone then N increases very rapidly and becomes of order of $10^{8}$. If we reduce the initial potential the order of the number of e-foldings also gets reduced. This fact is shown in the figure 1b.

Since inflation is hidden from the other sectors of the theory, it couples to lighter fields with gravitational strength of the order $\frac{\sqrt{8\pi }\mu}{M_{p}}$. Now for $\mu=2\times 10^{-8}$, we get $N=65$ for the initial inflaton field value, $\phi_{i}=0.094~M_{p}$ \cite{Bento2}. But as we got from the figure 1b ,i.e., a huge amount of efoldings can be obtained by larger values of $\phi_{i}$, e.g, $N=9.1\times 10^{6}$ for $\phi_{i}=0.5M_{p}$\cite{Bento2}.

\subsection{Hyperbolic potential}
Let $V=Mcosh m\phi$, where $M$ and $m$ are constants $m=\frac{\alpha\sqrt{8\pi}}{M_{p}}=\frac{2.22\times10^{-3}\sqrt{8\pi}}{M_{5}}$\cite{Paul1}
For this case
$$N=-\frac{4\pi M^{2A}}{\sigma M_{p}^{2} (2A+3) m^{2}}\times\left[-Hypergeometric2F1\left(\frac{3}{2}+A,1,\frac{5}{2}+A, \frac{V_{f}^{2}}{M^{2}}\right)\left(\frac{V_{f}}{M}\right)^{(3+2A)}\right]
$$
\begin{equation}\label{28}
-\frac{4\pi M^{2A}}{\sigma M_{p}^{2} (2A+3) m^{2}}\times\left[Hypergeometric2F1\left(\frac{3}{2}+A,1,\frac{5}{2}+A, \frac{V_{i}^{2}}{M^{2}}\right)\left(\frac{V_{i}}{M}\right)^{(3+2A)}\right]
\end{equation}

In figure 2a and 2b we have plotted the number of e-foldings during inflation with respect to $V_{f}$ and EoS parameter A ; and $V_{i}$ and EoS parameter A respectively. The curves are quite similar to figure 1a and 1b respectively.
\section{Inflation with polytropic fluid on the brane}
A polytropic process is a thermodynamic process that obeys the relation:
\begin{equation}\label{29}
p = {\cal D} \rho^{\gamma}
\end{equation}
where p is the pressure, $\rho$ is density, $\gamma$ is related to the the polytropic index,n, by the equation $\gamma=\frac{n+1}{n}$, where n is any real number. In the case of an isentropic ideal gas, $\gamma$ is the ratio of specific heats, known as the adiabatic index or as adiabatic exponent. ${\cal D}$ is a constant. This equation can be used to accurately characterize processes of certain systems, notably the compression or expansion of a gas and in some cases liquids and solids.

The equation is a valid characterization of a thermodynamic process assuming that the process is quasistatic. Under standard conditions, most gases can be accurately characterized by the ideal gas law. This construct allows for the pressure-volume relationship to be defined for essentially all ideal thermodynamic cycles, such as the well-known Carnot cycle. Note however that there may also be instances where a polytropic process occurs in a non-ideal gas.

For certain values of the polytropic index, n, the process will be synonymous with other common processes, like, when $n<0$, an explosion occurs. For $n=0$ we get a isobaric process whereas $n=1$ denotes a isothermal process. Now if $1<n<$ $\gamma$ then it indicates a quasi-adiabatic process such as in an internal combustion engine during expansion, or in vapor compression refrigeration during compression. Besides, $n=\gamma$ denotes a adiabatic process. Finally, $n=\infty$ denotes a isochoric process.

Using the equation (\ref{29}) in equation (\ref{17}) and integrating we get
\begin{equation}\label{30}
\rho^{*} = \left[-{\cal D} +\frac{\rho_{0}}{a^{3\left(1-\gamma\right)}}\right]^{\left(1-\gamma\right)}
\end{equation}
So the expression for $H^{2}$ here becomes
\begin{equation}\label{31}
H^{2}\simeq\kappa\left\{-{\cal D}+V^{(1-\gamma)}\right\}^{\frac{1}{(1-\gamma)}}\left[1+\frac{\left\{-{\cal D}+V^{(1-\gamma)}\right\}^{\frac{1}{(1-\gamma)}}}{2\sigma}\right]
\end{equation}
We also obtain
\begin{equation}\label{32}
\epsilon~~~\simeq ~~\frac{M_{p}^{2}}{16\pi}\left[\frac{V'^{2}V^{-\gamma}}{\left\{-{\cal D}+V^{(1-\gamma)}\right\}^{\frac{2-\gamma}{1-\gamma}}}\times \frac{\left[1+\frac{\left\{-{\cal D}+V^{(1-\gamma)}\right\}^{\frac{1}{(1-\gamma)}}}{\sigma}\right]}{\left[1+\frac{\left\{-{\cal D}+V^{(1-\gamma)}\right\}^{\frac{1}{(1-\gamma)}}}{2\sigma}\right]^{2}}\right]
\end{equation}
and
\begin{equation}\label{33}
\eta~~~\simeq ~~\frac{M_{p}^{2}}{8\pi}\frac{V''}{\left\{-{\cal D}+V^{(1-\gamma)}\right\}^{\frac{1}{(1-\gamma)}}}\times \left[1+\frac{\left\{-{\cal D}+V^{(1-\gamma)}\right\}^{\frac{1}{(1-\gamma)}}}{2\sigma}\right]^{-1}
\end{equation}
The condition for inflation is
\begin{equation}\label{34}
V^{-\gamma}V'^{2}\left[1+\frac{\left\{-{\cal D}+V^{(1-\gamma)}\right\}^{\frac{1}{(1-\gamma)}}}{\sigma}\right]<\frac{16\pi}{M_{p}^{2}}\left\{-{\cal D}+V^{(1-\gamma)}\right\}^{\frac{2-\gamma}{(1-\gamma)}}\left[1+\frac{\left\{-{\cal D}+\rho_{\phi}^{(1-\gamma)}\right\}^{\frac{1}{(1-\gamma)}}}{2\sigma}\right]^{2}
\end{equation}
Inflation ends when $\epsilon\simeq 1$ and we get
\begin{equation}\label{35}
V_{f}^{-\gamma}V_{f}^{'2}\left[1+\frac{\left\{-{\cal D}+V_{f}^{(1-\gamma)}\right\}^{\frac{1}{(1-\gamma)}}}{\sigma}\right]\approx\frac{16\pi}{M_{p}^{2}}\left\{-{\cal D}+V_{f}^{(1-\gamma)}\right\}^{\frac{2-\gamma}{(1-\gamma)}}\left[1+\frac{\left\{-{\cal D}+V_{f}^{(1-\gamma)}\right\}^{\frac{1}{(1-\gamma)}}}{2\sigma}\right]^{2}
\end{equation}
where subscript $'f'$ denotes the final value at the end of the inflation.

Now, in the high energy limit,
$$\left\{-{\cal D}+\rho_{\phi}^{(1-\gamma)}\right\}^{\frac{1}{(1-\gamma)}}\approx\left\{-{\cal D}+V^{(1-\gamma)}\right\}^{\frac{1}{(1-\gamma)}}\gg \sigma ,$$
 equation (\ref{35}) reduces to \cite{Herrera1}
$$V_{f}^{'2}\approx\frac{16\pi V_{f}^{\gamma}}{M_{p}^{2}}\left\{-{\cal D}+V^{(1-\gamma)}\right\}^{\frac{2-\gamma}{(1-\gamma)}}\left[1+\frac{\left\{-{\cal D}+V_{f}^{(1-\gamma)}\right\}^{\frac{1}{(1-\gamma)}}}{2\sigma}\right]^{2}\left[1+\frac{\left\{-{\cal D}+V_{f}^{(1-\gamma)}\right\}^{\frac{1}{(1-\gamma)}}}{\sigma}\right]^{-1}$$
\begin{equation}\label{36}
\approx \frac{4\pi V_{f}^{\gamma}}{M_{p}^{2}\sigma }\left\{-{\cal D}+V_{f}^{(1-\gamma)}\right\}^{\frac{3-\gamma}{1-\gamma}}
\end{equation}
The number of e-foldings in this case
\begin{equation}\label{37}
N=-\frac{8\pi}{M_{p}^{2}}\int_{V_{i}}^{V_{f}}\frac{1}{V'^{2}}\left\{-{\cal D}+V^{(1-\gamma)}\right\}^{\frac{1}{1-\gamma}}\left[1+\frac{\left\{-{\cal D}+V^{(1-\gamma)}\right\}^{\frac{1}{1-\gamma}}}{2\sigma}\right]dV
\end{equation}
In the high energy limit, equation (\ref{37}) becomes
\begin{equation}\label{38}
N=-\frac{4\pi}{\sigma M_{p}^{2}}\int_{V_{i}}^{V_{f}}\frac{\left\{-{\cal D}+V^{(1-\gamma)}\right\}^{\frac{2}{1-\gamma}}}{V'^{2}}dV
\end{equation}

For $V=\frac{1}{2}m^{2}\phi^{2}$,
$$N=-\frac{\pi}{4\sigma m^{2}M_{p}^{2}}[\left\{-2^{1-\gamma}{\cal D}+\left(2V_{f}\right)^{1-\gamma}\right\}^{\frac{2}{1-\gamma}}-\left\{-2^{1-\gamma}{\cal D}+\left(2V_{i}\right)^{1-\gamma}\right\}^{\frac{2}{1-\gamma}}+$$
$$\frac{2^{2-\gamma{\cal D}}}{1-\gamma}\{\left\{-2^{1-\gamma}{\cal D}+\left(2V_{f}\right)^{1-\gamma}\right\}^{\frac{2}{1-\gamma}}Hypergeometric2F1\left[\frac{2}{1-\gamma},1, \frac{3-\gamma}{1-\gamma},1-\frac{V_{f}^{1-\gamma}}{{\cal D}}\right]$$
\begin{equation}\label{38A}
-\left\{-2^{1-\gamma}{\cal D}+\left(2V_{i}\right)^{1-\gamma}\right\}^{\frac{2}{1-\gamma}}Hypergeometric2F1\left[\frac{2}{1-\gamma},1, \frac{3-\gamma}{1-\gamma},1-\frac{V_{i}^{1-\gamma}}{{\cal D}}\right]\}]
\end{equation}
In the low energy limit,
$\left\{-{\cal D}+V_{f}^{(1-\gamma)}\right\}^{\frac{1}{1-\gamma}}\ll \sigma$, so equation (\ref{35}) reduces to
\begin{equation}\label{39}
V_{f}^{'2}\approx\frac{16\pi V_{f}^{\gamma}}{M_{p}^{2}}\left\{-{\cal D}+V_{f}^{(1-\gamma)}\right\}^{\frac{2-\gamma}{1-\gamma}}
\end{equation}
and equaton (\ref{37}) remains the same.

In fig. 3a and 3b we have plotted the number of e-foldings $N$ with respect to the final and initial potentials respectively and $\gamma$, the EoS parameter. These two graphs are quite different than the graphs of the barotropic cases. At first we will observe the variation of $N$ w.r.t. $V_{f}$ and $\gamma$. Here we can see for constant $\gamma$, $N$ is not changing with respect to $V_{f}$. The main thing we can see is  that the smaller the value of $\gamma$, the graph is obtained at a lower value of $V_{f}$. But as we increase $\gamma$, the graph starts for large $V_{f}$. $N$ increases with increment of $\gamma$ but it reaches a maxima at a certain $\gamma=\gamma_{crit}$ then it reduces again.

We can interprete the above facts like this :- As pressure increases as the exponential order of density, so if the power $\gamma$ is high then the pressure is so high which decreases the number of e-foldings and the final potential increases as we increase $\gamma$. The rate of change of potential energy to kinetic energy (which actually increases the size of the universe) decreases. So the inflation is bit weaker for larger $\gamma$ which immediately leads to the decreasing of number of e-foldings.

Now we will analyse fig. 3b where change of N is observed with the change of $V_{i}$ in spite of $V_{f}$. At first we must tell that N increases with the increament of $V_{i}$. For constant $V_{i}$(large) as $\gamma$ increases the effect on $N$ resembles that of fig 3a. But $N$ exists for lower $V_{i}$ too. This not at all of surprize as $V_{i}$ is the initial potential which is to be choosen by us actually upon which $\gamma$ has no impact.
\section{Inflation with Modified Chaplygin Gas type fluid on the brane}

Within the framework of Einstein's gravity, due to this present accelerating phase, it is reasonable to believe  that DE is the dominating part ($74.5\%$ of the energy content in the observable universe) \cite{Bahcall} of the total energy of the universe. The candidates for DE \cite{Copeland,Alcaniz} can be classified as cosmological constant, dynamical component like quintessence, \cite{Wetterich, Ratra1, Caldwell, Ganzalez-Diaz, Fujii}, k-essence \cite{Armendariz-Picon1, Garriga,Chibd, Armendariz-Picon2, Armendariz-Picon3, Malquarti1, Malquarti2}, Chaplygin gas (CG) \cite{Kamenhchik, Gorini1, Gorini2, Alam, Chimento, Biesida, Bouhmadi-Lopez, Bento, Silva, Bertolami}. Although the cosmological constant is by far the simplest and the most popular candidate for DE, from the point of view of fine tuning and cosmic coincidence problem, it is not a suitable candidate for DE. On the other hand, dynamical DE models are favorable as they admit to construct `tracker' \cite{Zlatev, Steinhardt} or `attractor' \cite{Armendariz-Picon1,Takeshi} solutions. But most of the dynamical DE
models are described by a scalar field (often called quintessence field) which is unable to describe the transition
from a universe filled with matter to an exponentially expanding universe. However, the DE can be represented by
an exotic type of fluid known as CG \cite{Kamenhchik, Gorini1} having equation of state $p=-\frac{\beta}{\rho}$, where
$p$ and $\rho$ are respectively the pressure and energy density and $\beta$ is a positive constant.
Subsequently, this equation of state was generalized to $p=-\frac{\beta}{\rho^{n}},~ 0\leq n \leq 1$, and is known as generalized CG (GCG) \cite{Gorini1, Alam, Chimento, Biesida, Bouhmadi-Lopez, Bento, Silva, Bertolami}. Very recently a chaplygin gas model in the framework of Braneworld inflation using an exponential potential was studied \cite{Zarrouki}. In recent past, there was further modification to this equation of state as \cite{ Chimento,Benaoum, Debnath, Barreira}
\begin{equation}\label{40}
p=\alpha \rho - \frac{\beta}{\rho^{n}}
\end{equation}
and the model is known as modified CG (MCG). It gives the
cosmological evolution from an initial radiation era (with
$\alpha=\frac{1}{3}$) to (asymptotically) the $\Lambda CDM$ era
(where fluid behaves as cosmological constant). Based on
dimensionless age parameter ($H_{0}-t_{0}$) \cite{Dev1} and
observed $H(z)-z$ \cite{Wu1} data for both cold dark matter (CDM)
and unified dark matter energy (UDME) models, the values of
parameters $\alpha$ and $n$ are constrainted \cite{Thakur1}. In
order to obtain a viable cosmology with MCG, $\alpha$ should be
restricted to positive values.
%The constraints on $\alpha$ and $n$ are :
%(i) $0\leq\alpha\leq 1.07$  and $0\leq n\geq 1$  in CDM
%and
%(ii)$0\leq\alpha\leq 1.35$  and $0\leq n\geq 1$  in UDME.
Besides this, there are best-fit values of the parameters for CDM and UDME models which correspond to
$H_0-t_0$ data. The permissible values of the parameters are given in the following table.

\begin{center}
{\bf Table I:} Permissible ranges of MCG parameters in different models.\\
\begin{tabular}{|l|}
\hline\hline
~~Source of data
~~~~~~~~~~~~~~~~~~~~~~~~~~~~~~~~~~~~~~~~~~~~~~~~~~~~~~~~Type of Model \\ \hline
\\
~~~~~~~~~~~~~~~~~~~~~~~~~~~~~~~~~~~~~~~~~~~~~~~~~~~~~~~~~~~~~~CDM~~~~~~~~~~~~~~~~~~~~~~~~~~~~~~~~UDME~~~~~~~
\\\hline
~~~~~~~~~~~~~~~~~~~~~~~~~~~~~~~~~~~~~~~~~~~~~~~~~~~~~~$\alpha$~~~~~~~~~~~~~~~~~~$n$~~~~~~~~~~~~~~~~~~~~$\alpha$~~~~~~~~~~~~~~~~~~$n$~~~~~~~
\\\hline\hline
\\
~$H(z)-z$~~~~~~~~~~~~~~~~~~~~~~~~~~~~~~~~~$0\leq\alpha\leq 1.07$~~~~~~$0\leq n\leq 1$~~~~~~~$0\leq\alpha\leq 1.35$~~~~~$0\leq n\leq 1$
\\\\\hline\\
~$H_{0}-t_{0}$~~~~~~~~~~~~~~~~~~~~~~~~~~~~~~~~~~~~~~$\alpha=0.01$~~~~~~~~~$ n=0.01$~~~~~~~~~~$\alpha=0.06$~~~~~~~~~$ n=0.11$~\\
(best-fit)
\\\\ \hline\hline
\end{tabular}
\end{center}
So, using  EoS (\ref{40}) we have
\begin{equation}\label{40A}
\rho^{*}=\left[\frac{1}{\alpha+1}\left\{\beta+\frac{\rho_{0}}{a^{3(\alpha+1)(n+1)}}\right\}\right]^{\frac{1}{n+1}}
\end{equation}
Now the expression for $H^{2}$ is
\begin{equation}\label{41}
H^{2}\simeq \kappa\left\{\frac{1}{\alpha+1}\left(\beta+V^{\left(\alpha+1\right)\left(n+1\right)}\right)\right\}^{\frac{1}{n+1}}\left[1+\frac{\left\{\frac{1}{\alpha+1}\left(\beta+V^{\left(\alpha+1\right)\left(n+1\right)}\right)\right\}^{\frac{1}{n+1}}}{2\sigma}\right]
\end{equation}
and the slow roll parameters are
\begin{equation}\label{42}
\epsilon\simeq\frac{M_{p}^{2}}{16\pi}\left[\frac{V^{\alpha+n(\alpha+1)}V'^{2}}{\left\{\frac{1}{\alpha+1}\left(\beta+V^{\left(\alpha+1\right)\left(n+1\right)}\right)\right\}^{\frac{n+2}{n+1}}}\times\frac{\left\{1+\frac{\left\{\frac{1}{\alpha+1}\left(\beta+V^{\left(\alpha+1\right)\left(n+1\right)}\right)\right\}^{\frac{1}{n+1}}}{\sigma}\right\}}{\left\{1+\frac{\left\{\frac{1}{\alpha+1}\left(\beta+V^{\left(\alpha+1\right)\left(n+1\right)}\right)\right\}^{\frac{1}{n+1}}}{2\sigma}\right\}^{2}}\right]
\end{equation}
and
\begin{equation}\label{43}
\eta\simeq\frac{M_{p}^{2}}{8\pi}\times\frac{V''}{\left\{\frac{1}{\alpha+1}\left(\beta+V^{\left(\alpha+1\right)\left(n+1\right)}\right)\right\}^{\frac{1}{n+1}}}\times\left\{1+\frac{\left\{\frac{1}{\alpha+1}\left(\beta+V^{\left(\alpha+1\right)\left(n+1\right)}\right)\right\}^{\frac{1}{n+1}}}{2\sigma}\right\}^{-1}
\end{equation}
Then condition for inflation gives the constraint upon $V$ as,
$$V^{\alpha+n(\alpha+1)}V'^{2}\left\{1+\frac{\left\{\frac{1}{\alpha+1}\left(\beta+V^{\left(\alpha+1\right)\left(n+1\right)}\right)\right\}^{\frac{1}{n+1}}}{\sigma}\right\}~~~~~~~~~~~~~~~~~~~~~~~~~~~~~~~~~~~~~~~~~~$$
\begin{equation}\label{44}
~~~~~~~~~<\frac{16\pi}{M_{p}^{2}}\left\{\frac{1}{\alpha+1}\left(\beta+V^{\left(\alpha+1\right)\left(n+1\right)}\right)\right\}^{\frac{n+2}{n+1}}\left\{1+\frac{\left\{\frac{1}{\alpha+1}\left(\beta+V^{\left(\alpha+1\right)\left(n+1\right)}\right)\right\}^{\frac{1}{n+1}}}{2\sigma}\right\}^{2}
\end{equation}
Inflation ends when
$$V_{f}^{\alpha+n(\alpha+1)}V_{f}'^{2}\left\{1+\frac{\left\{\frac{1}{\alpha+1}\left(\beta+V_{f}^{\left(\alpha+1\right)\left(n+1\right)}\right)\right\}^{\frac{1}{n+1}}}{\sigma}\right\}~~~~~~~~~~~~~~~~~~~~~~~~~~~~~~~~~~~~~~~~~~$$
\begin{equation}\label{45}
~~~~~~~~~\simeq\frac{16\pi}{M_{p}^{2}}\left\{\frac{1}{\alpha+1}\left(\beta+V_{f}^{\left(\alpha+1\right)\left(n+1\right)}\right)\right\}^{\frac{n+2}{n+1}}\left\{1+\frac{\left\{\frac{1}{\alpha+1}\left(\beta+V_{f}^{\left(\alpha+1\right)\left(n+1\right)}\right)\right\}^{\frac{1}{n+1}}}{2\sigma}\right\}^{2}
\end{equation}
where $V_{f}$ is the value of the potential at the end of the inflation.

In the high energy limit when $\left\{\frac{1}{\alpha+1}\left(\beta+V^{\left(\alpha+1\right)\left(n+1\right)}\right)\right\}^{\frac{1}{n+1}}\gg \sigma$ equation(\ref{45}) reduces to
\begin{equation}\label{46}
V_{f}'^{2}\simeq\frac{4\pi}{M_{p}^{2}\sigma}\frac{\left\{\frac{1}{\alpha+1}\left(\beta+V_{f}^{\left(\alpha+1\right)\left(n+1\right)}\right)\right\}^{\frac{n+3}{n+1}}}{V_{f}^{\alpha+n(\alpha+1)}}
\end{equation}
The number of e-folds during the inflation is given by,
\begin{equation}\label{47}
N=-\frac{8\pi}{M_{p}^{2}}\int_{V_{i}}^{V_{f}}\frac{1}{V'^{2}}\left\{\frac{1}{\alpha+1}\left(\beta+V^{\left(\alpha+1\right)\left(n+1\right)}\right)\right\}^{\frac{1}{n+1}}\left[1+\frac{\left\{\frac{1}{\alpha+1}\left(\beta+V^{\left(\alpha+1\right)\left(n+1\right)}\right)\right\}^{\frac{1}{n+1}}}{2\sigma}\right]dV
\end{equation}
In the high energy limit equation (\ref{47}) becomes
\begin{equation}\label{48}
N=-\frac{4\pi}{M_{p}^{2}\sigma}\int_{V_{i}}^{V_{f}}\frac{1}{V'^{2}}\left\{\frac{1}{\alpha+1}\left(\beta+V^{\left(\alpha+1\right)\left(n+1\right)}\right)\right\}^{\frac{2}{n+1}}dV
\end{equation}
The subscripts 'i' and 'f' are used to denote the initial value and the final value at the beginning and the end of the inflation respectively.

And in the low energy limit when $\left\{\frac{1}{\alpha+1}\left(\beta+V^{\left(\alpha+1\right)\left(n+1\right)}\right)\right\}^{\frac{1}{n+1}}\ll \sigma$ equation(\ref{45}) reduces to
\begin{equation}\label{49}
V_{f}^{'2}\simeq\frac{16\pi}{M_{p}^{2}}\frac{\left\{\frac{1}{\alpha+1}\left(\beta+V_{f}^{\left(\alpha+1\right)\left(n+1\right)}\right)\right\}^{\frac{n+2}{n+1}}}{V_{f}^{\alpha+n(\alpha+1)}}
\end{equation}
and the equation (\ref{47}) remains the same.
Now if the form of the potential be $\frac{1}{2}m^{2}\phi^{2}$
then we have the expression for number of the efolds as
$$N=\frac{4^{-2-\alpha}\pi (1+\alpha)^{-\frac{3+n}{1+n}}}{m^{2}\sigma M_{p}^{2}}\left\{2^{(1+n)(1+\alpha)}\left(\beta+V_{f}^{(1+n)(1+\alpha)}\right)\right\}^{\frac{2}{1+n}}\times
$$$$\left\{\beta Hypergeometric2F1\left(\frac{2}{1+n}, 1, \frac{3+n}{1+n}, \beta+V_{f}^{(1+n)(1+\alpha)}\right)-1\right\}$$
$$-\frac{4^{-2-\alpha}\pi
(1+\alpha)^{-\frac{3+n}{1+n}}}{m^{2}\sigma
M_{p}^{2}}\left\{2^{(1+n)(1+\alpha)}\left(\beta+V_{i}^{(1+n)(1+\alpha)}\right)\right\}^{\frac{2}{1+n}}\times$$
\begin{equation}\label{50}
\left\{\beta Hypergeometric2F1\left(\frac{2}{1+n}, 1,
\frac{3+n}{1+n}, \beta +V_{i}^{(1+n)(1+\alpha))}\right)-1\right\}
\end{equation}

In figure 4(a) and 4(b) again we plot the number of
e-foldings with respect to potential (final and initial
respectively) and MCG parameter n. Here for constant $n$, $V_{f}$
has no impact on $N$. But as we increase $n$ we can see that after
a certain value of n, N does not increase with it. But no slow
roll takes place as well. That is if n is sufficiently large rate
of increasing number of e-foldings,i.e., $\frac{\partial
N}{\partial n}$ will be small than the lower n cases. We can
speculate that a big n indicates a high negative pressure in
equation (\ref{40}) which may cause a Big Rip in such a way that
inflation may not have a physical meaning beyond that. So N does
not increase any more. In 4b we can see that if we increase
$V_{i}$ at a lower range of $\gamma$ then we can see the number of
e-foldings increases sharply. This is quite similar as the cases
$1b,2b~and~3b$.

\section{Inflation with fluid obeying Linear red shift parametrization(LRP) of the EoS parameter}
If we start with the assumption $\omega$=constant, the observational data favors $\omega=-1$. But in case of dynamical DE, the current observational data does not yield any conclusive constraint over the equation of state parameter $\omega(z)$. The red shift of the supernova $z=a^{-1}-1$ measures the scale factor $a$, the size of the universe when the supernova exploded relative to its current size. The astrophysical constraints discussed below arise from well established cosmological observations at different red shifts.

{\bf (A) LSS Constraint : \\}
During the galaxy formation era $(1 < z < 3)$, dark energy density must be subdominant to matter density; accordingly $\Omega_{X} < 0.5$ \cite{Freedman1}.

{\bf (B) CMB Constraint : \\}
Caldwell \cite{Caldwell1} and Upadhye et al. \cite{Upadhye1} have argued that $\left(\Omega_{X}\right)-{dec} < 0.1$ at $z = 1100$. It
imposes a high redshift upper bound ΩX for any viable parametrization of the equation of
state parameter $\omega_(z)$.

{\bf(C) BBN Constraint : \\}
The presence of dark energy until nucleosynthesis epoch, should not disturb the observed
Helium abundance in the universe which is regarded as the foundation stone of the Big Bang
Theory. According to Johri \cite{Johri1} $\left(\Omega_{X} \right)_{BBN} < 0.14$ at $z = 10^{10}$ whereas the latest analysis
of Cyburt \cite{Cyburt1} constrains $\left(\Omega_{X} \right)_{BBN} < 0.21$ at $z = 10^{10}$ . This puts a stringent high redshift
limit on dark energy.
The equation of state parameter w(z) is given by Huterer et al. \cite{Huterer1} and Weller et al.
\cite{Weller1} as
\begin{equation}\label{51}
\omega=\omega_{0}+\omega_{1}z
\end{equation}
It has been used by Riess et al. \cite{Riess1} for probing SNIa observations at $z < 1$. The best fit
values to SNIa gold set data \cite{Dicus1} are $\omega_{0} = 1.4$, $\omega_{1} = 1.67$.

Hence, the parameters favor dark energy of phantom origin. In the distant past $( z\gg 1)$, $\omega \rightarrow\infty$ and in the distant future $(1 + z \rightarrow 0)$, $\omega=\omega_{0}-\omega_{1}=-3.07$.

We note that
$$(i) a\rightarrow 0, i.e., z\rightarrow \infty, \rho\rightarrow\infty, p\rightarrow\infty ~~~~~~~~~~~~~~~~~$$
$$(ii) a\rightarrow \infty, i.e., z\rightarrow -1 ~:~~~~~~~~~~~~~~~~~~~~~~~~~~~~~~~~~$$
$$(a)\rho\rightarrow 0, ~p\rightarrow 0~if~ \omega_{0}>\omega_{1}~~~~~~~~~~~~~~~~~ $$
$$(b)\rho\rightarrow \infty, ~p\rightarrow -\infty~if~ \omega_{0}<\omega_{1}<1+\omega_{0} $$
$$~~~~~~~~~~~~~~~~~~~~~~~~~~~~(c)\rho\rightarrow \rho_{0}exp\left\{-3\omega_{1}\right\}~ and ~p\rightarrow -\rho_{0}exp\left\{-3\omega_{1}\right\}~if~ \omega_{1}=1+\omega_{0} $$
The limiting behavior shows that for realistic fluid we must have $\omega_{0}>\omega_{1}$. Here, the energy density has the correct behavior, pressure is positive throughout the evolution (approaching zero) with a finite sound velocity. When $\omega_{0}<\omega_{1}$, the pressure becomes negative at an intermediate stage of the evolution and finally approaches zero -a possibility of an accelerating phase of the universe. For $\omega_{0}=\omega_{1}$,the pressure cannot be negative at any stage of the evolution \cite{Mazumder1}.

From the energy conservation equation we have
\begin{equation}\label{51A}
\rho(z)=\rho_{0}\left(1+z\right)^{3\left(1+\omega_{0}-\omega_{1}\right)}exp\left\{3\omega_{1}z\right\}
\end{equation}
Thus,
\begin{equation}\label{52}
\rho^{*}=\rho_{\phi}^{\left(1+\omega_{0}-\omega_{1}\right)}exp\left\{3\omega_{1}\left(\rho_{\phi}^{\frac{1}{3}}-1\right)\right\}
\end{equation}
The expression for $H^{2}$ is
\begin{equation}\label{53}
H^{2}=\kappa\left[ exp\left\{3\omega_{1}\left(V^{\frac{1}{3}}-1\right)\right\}V^{\left(1+\omega_{0}-\omega_{1}\right)}\right]\left[1+\frac{exp\left\{3\omega_{1}\left(V^{\frac{1}{3}}-1\right)\right\}V^{\left(1+\omega_{0}-\omega_{1}\right)}}{2\sigma}\right]
\end{equation}
and the slow roll parameters are
$$\epsilon=\frac{M_{p}^{2}}{8\sqrt{2}\pi \sigma}\times exp\{6\omega_{1}(V^{\frac{1}{3}}-1)\}\{1+\omega_{0}+\omega_{1}
(-1+V^{\frac{1}{3}})\}V^{\omega_{0}-2\omega_{1}}(V^{1+\omega_{0}}+\sigma
V^{\omega_{1}}exp\{-3\omega_{1}(V^{\frac{1}{3}}-1)\})V'^{2}$$
$$\times\{exp\{3\omega_{1}(V^{\frac{1}{3}}-1)\}V^{1+\omega_{0}-2\omega_{1}}(V^{1+\omega_{0}}exp\{3\omega_{1}(V^{\frac{1}{3}-1})\}+2\sigma
V^{\omega_{1})}\}^{-\frac{1}{2}}$$
\begin{equation}\label{54}
\times\{V^{(1+\omega_{0}-\omega_{1})}exp\{3\omega_{1}(V^{\frac{1}{3}}-1)\}
+\frac{V^{2(1+\omega_{0}-\omega_{1})}exp\{-6\omega_{1}(1-V^{\frac{1}{3}})\}}{2\sigma}\}^{-\frac{3}{2}}
\end{equation}
\begin{equation}\label{55}
\eta=\frac{M_{p}^{2}}{8\pi}\frac{V''}{\left[V^{\left(1+\omega_{0}-\omega_{1}\right)}exp\left\{-3\omega_{1}\left(V^{\frac{1}{3}}-1\right)\right\}\right] \left[1+\frac{V^{\left(1+\omega_{0}-\omega_{1}\right)}exp\left\{3\omega_{1}\left(1-V^{\frac{1}{3}}\right)\right\}}{2\sigma}\right]}
\end{equation}
Then condition for inflation gives the constraint upon $V$ as
$$exp\left\{6\omega_{1}\left(V^{\frac{1}{3}}-1\right)\right\}\left\{1+\omega_{0}+\omega_{1}\left(-1+V^{\frac{1}{3}}\right)\right\}V^{\omega_{0}-2\omega_{1}}\left(V^{1+\omega_{0}}+\sigma V^{\omega_{1}}exp\left\{-3\omega_{1}\left(V^{\frac{1}{3}}-1\right)\right\}\right)V'^{2}\leq$$
$$\frac{8\sqrt{2}\pi \sigma}{M_{p}^{2}}\times\left\{V^{\left(1+\omega_{0}-\omega_{1}\right)}exp\left\{3\omega_{1}\left(V^{\frac{1}{3}}-1\right)\right\}    +\frac{V^{2\left(1+\omega_{0}-\omega_{1}\right)}exp\left\{-6\omega_{1}\left(1-V^{\frac{1}{3}}\right)\right\}}{2\sigma}\right\}^{\frac{3}{2}}\times$$
\begin{equation}\label{55A}
\sqrt{exp\left\{3\omega_{1}\left(V^{\frac{1}{3}}-1\right)\right\}V^{1+\omega_{0}-2\omega_{1}}\left(V^{1+\omega_{0}}exp\left\{3\omega_{1}
\left(V^{\frac{1}{3}-1}\right)\right\}+2\sigma
V^{\omega_{1}}\right)}
\end{equation}
Inflation ends when
$$exp\left\{6\omega_{1}\left(V_{f}^{\frac{1}{3}}-1\right)\right\}\left\{1+\omega_{0}+\omega_{1}\left(-1+V_{f}^{\frac{1}{3}}\right)\right\}V_{f}^{\omega_{0}-2\omega_{1}}\left(V_{f}^{1+\omega_{0}}+\sigma V_{f}^{\omega_{1}}exp\left\{-3\omega_{1}\left(V_{f}^{\frac{1}{3}}-1\right)\right\}\right)V_{f}'^{2}\simeq$$
$$\frac{8\sqrt{2}\pi \sigma}{M_{p}^{2}}\times\left\{V_{f}^{\left(1+\omega_{0}-\omega_{1}\right)}exp\left\{3\omega_{1}\left(V_{f}^{\frac{1}{3}}-1\right)\right\}    +\frac{V_{f}^{2\left(1+\omega_{0}-\omega_{1}\right)}exp\left\{-6\omega_{1}\left(1-(V_{f}^{\frac{1}{3}}\right)\right\}}{2\sigma}\right\}^{\frac{3}{2}}\times$$
\begin{equation}\label{55B}
\sqrt{exp\left\{3\omega_{1}\left(V_{f}^{\frac{1}{3}}-1\right)\right\}V_{f}^{1+\omega_{0}-2\omega_{1}}\left(V_{f}^{1+\omega_{0}}
exp\left\{3\omega_{1}\left(V_{f}^{\frac{1}{3}-1}\right)\right\}+2\sigma
V_{f}^{\omega_{1}}\right)}
\end{equation}
where $V_{f}$ is the value of the potential at the end of the inflation.

The number of e-folds is given by
\begin{equation}\label{56}
N=-\frac{8\pi}{M_{p}^{2}}\int_{V_{i}}^{V_{f}}\frac{V^{(1+\omega_{0}-\omega_{1})}exp\left\{3\omega_{1}\left(V^{\frac{1}{3}}-1\right)\right\}}{V'^{2}}\left[1+\frac{V^{\left(1+\omega_{0}-\omega_{1}\right)}exp\left\{3\omega_{1}\left(V^{\frac{1}{3}}-1\right)\right\}}{2\sigma}\right]dV
\end{equation}
Where the subscripts 'i' and 'f' denote initial and final values of the potential at the beginning and the end of the inflation respectively.

In the high energy limit when $V^{\left(1+\omega_{0}-\omega_{1}\right)}exp\left\{3\omega_{1}\left(V^{\frac{1}{3}}-1\right)\right\}\gg\sigma$
\begin{equation}\label{56A}
N=-\frac{4\pi}{\sigma M_{p}^{2}}\int^{V_{f}}_{V_{i}}\frac{V^{2\left(1+\omega_{0}-\omega_{1}\right)}exp\left\{6\omega_{1}\left(V^{\frac{1}{3}}-1\right)\right\}}{V^{'2}}dV
\end{equation}
If the potential is of the form $\frac{1}{2}m^{2}\phi^{2}$ then
\begin{equation}\label{57}
N=-\frac{6\pi}{\sigma m^{2} M_{p}^{2}\left(5+6\omega_{0}\right)}\left[exp\left\{\left(V_{f}^{\frac{1}{3}}-1\right)\left(5+6\omega_{0}\right)\right\}-exp\left\{\left(V_{i}^{\frac{1}{3}}-1\right)\left(5+6\omega_{0}\right)\right\}\right]
\end{equation}
where we have assumed $\omega_{0}=\omega_{1}-\frac{5}{6}$.

Figure (5a) and (5b) are quite similar with the (1a) and (1b)
respectively. Now if we look upon the EOS $p=({\omega}_{0}
+{\omega}_{1} z)\rho$ and $p=A \rho$ and compare them, we can see
from some observational data \cite{Liu1} we can predict that
${\omega}_{0}=-1.39978 \pm 0.249302$ and ${\omega}_{1} =1.66605
\pm 0.892594$. Now if we take all the lower cases and compare
$\omega(z)$ with $A=-1~to~+1$ then we can see the corresponding
$z$ range will be 0.839197 to 3.42499 which is very much similar
to the range of $z$ during the galaxy formation era (LSS
constraint) \cite{Bertolami}.

\section{Discussions and Conclusion}
In this paper we have studied the inflation in brane world scenario for different kinds of fluids which fill up the universe. When we assume the cosmological constant $\Lambda_{4}$ to vanish, we can have $H^{2}$ as  a function of the matter confined in the brane and the brane tension using which we can obtain many information regarding inflation. In each case, we have taken a particular form of potential namely chaotic potential. We have given stress upon the number of e-foldings. This number represents the multiplication of the size of the universe at the end of inflation. At first while for barotropic case we have plotted the $N$-curve , we have checked the dependency of $N$ upon $V_{f}$, $V_{i}$, the final and initial potentials of the universe during inflation and the barotropic EoS parameter, $A$. The domain of $A$ is taken in such a way that it satisfies the whole range of phantom barrier to stiff fluid. It is observed that as we move from -1 to +1, the number of e-foldings increases except for the case when  $V_{i}$ is very low (to produce enough kinetic energy to give larger inflation, or larger N). While choosing the domains of final potential have kept it in mind that inflation increases the entropy, the dynamical movements in universe, i.e., the potential energy at the starting of the inflation converts into kinetic energy to support the expansion of the universe, now it is observed that whatever be the value of the final potential the value of N does not change with that which is obvious as we get $V_{f}$ at the end of inflation which of course has no such impact upon any physical quantity that is determined through the inflationary process. Besides, if potential energy is considered to be larger at the beginning of inflation then it is very obvious to get more kinetic energy converted from it which will amplify the inflationary process and thus the number of e-foldings will also have a larger value. For barotropic case only we have taken hyperbolic potential also to explain the inflation. The graph of $N$ resembles with that for the SSP we have taken the multiplicative factor M to be unity for which $\cosh m\phi(=1+\frac{1}{2}m^{2}\phi^{2}+\frac{1}{4!}m^{4}\phi^{4}+.......)$ is kind of an infinite power series of SSP. So it is very natural to get similar result as SSP. Whatever be the values of N drawn from graph they are nearly equal to the values obtained theoretically from literature \cite{Bento}. In  section 3 we have calculated all the data required to analyse the inflation of the universe filled with fluid of polytropic nature. Here again we have plotted the graphs of N as before the dependencies upon $V_{f}$ and $V_{i}$ are almost same as before. But here when we vary $\gamma$ over its physical range we find a point $\gamma_{crit}$ on $\gamma$-axis such that if it has been crossed the inflation ends with a smaller N than at that particular $\gamma_{crit}$. To explain this phenomena we can speculate that after growing enough large, $\gamma$ makes positive pressure so high that it rapidly ends the inflation such that the size of the universe can not be
larger than a particular value say $N(\gamma_{crit})$. If we can make $\gamma$ larger than $\gamma_{crit}$ then the large positive pressure stops inflation before its $N$ reaches $N(\gamma_{crit})$. In section 4 we have taken MCG and derived different physical quantities like slow roll parameter and $N$. We have plotted the graphs of $N$ against $V_{f}~and~V_{i}$ and the MCG parameter n (with the domain containing the interval which best fits to explain some observational evidences ). MCG provides negative pressure so inflation has not been stopped by it at a particular end (like polytropic case)but after we take a large $N$ we see that the rate of increment of $N$ is very slow(though not negative.). It is very obvious for a negative pressure generating gas like MCG . One fact is yet to be noted in barotropic, polytropic and MCG case that the order of N matches with calculations \cite{Bento} too. In the section 5 as before all the inflational quantities have been derived and at last the plots of N are given which are quite similar with barotropic case except the fact that the value of N is much much higher than the previous cases.

So at the end we can conclude that the natures of different parameters related to inflation are quite similar if we take the fluid present in the brane as of barotropic and linear red shift parametrization as almost in both cases pressure is positive with the variation of density $\rho$. But when we have considered MCG or polytropic fluid due to their negative or highly  positive pressure creating property the number of e-folds in MCG and polytropic are quite different than barotropic and LRP case.

As there is a severe impact upon the number of e-foldings when we increase the potential, so which form of the potential will be discussed is of most relevance. For barotropic case , we have taken two different potentials chaotic and hyperbolic and observed that the results are identical. Yet, there are different potentials like quartic potential and coulombic potential for which how the value of the number of e foldings changes will be matter of interest for an future work. We would also like to consider for our future work the cases in which the universe is filled with a perfect fluid having the equation of state $p=\omega(z)\rho$ where $\omega(z)$  can be taken as (i) Jassal Bagla Padmanabhan parametrization $\omega(z)=\omega_{0}+\frac{\omega_{1}z}{\left(1+z\right)^{2}}$, (ii) $\omega(z)=\omega_{0}+\frac{\omega_{1}z}{\left(1+z\right)}$ and $\omega=\omega_{0}+\omega_{1}z$ if $z<1$, $\omega_{0}+\omega_{1}$ if $z\geq 1$.

{\bf Acknowledgement :}\\
SM thanks Jadavpur University for allowing her to use the library and laboratory facilities. RB and NM thank West Bengal State Govt. and CSIR, India for awarding JRF.
\\

\end{document}